\documentstyle[preprint,aps]{revtex}
\begin{document}
\preprint{\font\fortssbx=cmssbx10 scaled \magstep2
%\hskip.5in \raise.1in\hbox{\fortssbx University of Wisconsin - Madison}
\hfill$\vcenter{\hbox{\bf MAD/PH/764}
                \hbox{\bf IFT-P.031/93}
                \hbox{\bf IFUSP-P 1047}
                \hbox{June 1993}
                \hbox{hep-ph/9306306}}$}
%\vspace{.2in}
%
\title{Quartic Anomalous Couplings in {\boldmath $e\gamma$} Colliders}
\author{O.\ J.\ P.\ \'Eboli\cite{pa1}} 
\address{Instituto de F\'{\i}sica,  Universidade de S\~ao Paulo, \\
Caixa Postal 20516,  CEP 01498-970 S\~ao Paulo, Brazil.}
\author{M.\ C.\ Gonzal\'ez-Garc\'{\i}a\cite{pa2}, and 
 S.\ F.\ Novaes\cite{pa3}}
\address{Physics Department, University of Wisconsin, \\
Madison, WI 53706,  USA.}
\maketitle
\thispagestyle{empty}

\begin{abstract}
We study the production of gauge boson pairs at the next
generation of linear $e^+e^-$ colliders operating in the
$e\gamma$ mode. The processes $e\gamma \rightarrow VV^\prime
F$ ($V,V^\prime =W$, $Z$, or $\gamma$ and $F=e$ or $\nu$)
can give valuable information  on  possible deviations  of
the quartic vector boson couplings from the Standard Model
predictions. We establish the range of the new couplings
that can be explored in these colliders based on a $3\sigma$
effect in the total cross section. We also present several
kinematical distributions of the final state particles that
could manifest the underlying new dynamics. Our results show
that an $e\gamma$ collider can extend considerably the
bounds on anomalous interactions coming from oblique
radiative corrections and from direct searches in $e^+e^-$
colliders.
\end{abstract}
\pacs{}

\newpage

%******************************************************************************
\section{Introduction}

The Standard Model (SM) of electroweak interactions has so
far explained all the available experimental data extremely
well \cite{sm}.  From LEP data, we can witness the agreement
between theory and experiment at the level of 1\%, which is
a striking confirmation of the $SU(2)_L\times U(1)_Y$
invariant interactions involving fermions and gauge bosons.
However, some predictions of the SM have not yet been object
of direct experimental observation, such as the symmetry
breaking mechanism and the interaction among the gauge
bosons. In particular, the structure of the trilinear and
quartic vector boson couplings is completely determined by
the non-abelian gauge structure of the model, and a detailed
study of these interactions can either confirm  the local
gauge  invariance of the theory or indicate the existence of
new physics beyond the SM. 

The triple and quartic gauge-boson vertices, in the scope of
gauge theories, have a common origin and possess a universal
strength. In the SM they are constrained by the  $SU(2)_L
\times U(1)_Y$ gauge invariance, and for a given process,
there are important cancellations between the bad high
energy behaviour of the various amplitudes, which leads to a
final result that is consistent with the unitarity
requirements.

In a more general context, the quartic anomalous couplings
can be related to the low energy limit of heavy state
exchange giving rise to an effective contact interactions,
whereas trilinear couplings  are modified by
integrating out heavy fields. Therefore,  it is possible to
build extensions of the SM where the trilinear couplings
remain equal to the SM ones, while the quartic vertices are
modified. For instance, the introduction of a new heavy
scalar singlet, that interacts strongly with the Higgs
sector of the SM, enhances the four vector-boson
interaction, without affecting neither the triple vector
boson coupling  nor spoiling the SM predictions for the
$\rho$ parameter \cite{hill}. Further examples include the
Higgs-less models \cite{bess} where the breaking of the
$SU(2)_L\times U(1)_Y$ symmetry is realized nonlinearly,
leading  to the appearance of large anomalous four gauge
boson vertices.

One of the main goals of LEP II at CERN will be the study of the
reaction $e^+e^- \rightarrow W^+W^-$ which will furnish 
bounds on possible anomalous $W^+W^-\gamma$ and
$W^+W^-Z$ interactions \cite{zenpa}.  Within the SM
framework, it will be also possible to constrain indirectly
the four-vector-boson interactions since the $SU(2)_L \times
U(1)_Y$ local gauge symmetry connects trilinear and quartic
vertices. Nevertheless, the test of a wide class of models,
that  predicts only quartic anomalous vector boson
interactions, will have to rely on the direct measurements
of reactions such as  $e^+e^- \rightarrow VVV$, $\gamma +
\gamma \rightarrow V V$, or $e + \gamma \rightarrow
VV^\prime F$ ($V, V^\prime =W$, $Z$, or $\gamma$ and $F=e$ or
$\nu$).  This, of course, cannot be done at LEP II due to
its limited available center of  mass energy. 

The planned next generation of linear $e^+e^-$ colliders
(NLC) will have enough center of mass energy to study
directly quartic vertices through the production of three
gauge boson \cite{cana1,barger,niaki}, and will offer
further independent tests of the SM. Moreover, at the NLC it
will be possible to transform an electron beam into a photon
one through the laser backscattering mechanism
\cite{las0,telnov}.  This kind of process will allow the NLC
to operate  in three different modes: $e^+e^-$, $e\gamma$,
and $\gamma\gamma$.  This versatility opens up the
opportunity to perform a deeper analysis of the anomalous
couplings, mainly for those new interactions involving
photons.  In fact, the study of the anomalous vertices
$W^+W^-\gamma\gamma$ and $Z^0Z^0\gamma\gamma$ has been
carried out in Ref.\ \cite{cana2} assuming the
$\gamma\gamma$ mode for NLC, where the sensitivity to
anomalous couplings is improved by two orders of magnitude
when compared to the $e^+e^-$ mode results. 

It is important to point out that the laser backscattering
mechanism leads to the total loss of the electron (positron)
beam in an $e^+e^-$ linear collider. Consequently, the
experimental collaborations of the NLC will have to decide
in which mode they want to operate the collider, {\em
i.e.\/} either in the $e^+e^-$, $e\gamma$ or $\gamma\gamma$
mode, and it is essential to study comparatively the
capabilities of each of these setups to explore new physics.

In this work we study the capability of an $e^+e^-$ linear
collider, operating in the $e\gamma$ mode, to search for
anomalous four-gauge-boson interactions. We restrict
ourselves to the analyses of  effective lagrangians that
contain at least one photon and that do not exhibit a
trilinear coupling associated to it since, in this case, they
cannot be indirectly probed at LEP II. This search for anomalous
couplings is carried out through the study of the reactions 
\begin{mathletters}
\begin{equation}
e + \gamma \rightarrow W^+ + W^- + e        \;\;\;\;\;\;\;\;  [WWE] \; ,
\label{wwe}  
\end{equation}
\begin{equation}                                             
e + \gamma  \rightarrow Z^0 + \gamma + e     \;\;\;\;\;\;\;\;  [ZGE] \; ,
\label{zge}
\end{equation}
\begin{equation}
e + \gamma  \rightarrow Z^0 + Z^0 + e        \;\;\;\;\;\;\;\;  [ZZE] \; ,
\label{zze}
\end{equation}
\begin{equation}
e + \gamma \rightarrow W^- + \gamma + \nu   \;\;\;\;\;\;\;\;  [WGN]\; ,
\label{wgn}
\end{equation}
\begin{equation}
e + \gamma \rightarrow W^- + Z^0 + \nu      \;\;\;\;\;\;\;\;  [WZN]\; ,
\label{wzn}
\end{equation}
%\label{pro}
\end{mathletters}
that receive contributions from the quartic vertices
$W^+W^-\gamma\gamma$, $W^+W^-Z^0\gamma$, and $Z^0Z^0\gamma
\gamma$.  Some of these reactions, assuming SM couplings for
all particles, were recently considered in Ref.\
\cite{che:vv}.  

The outline of this paper is as follows. In Sec.\
\ref{coupsec} we exhibit the chosen effective operators that
gives rise to the anomalous couplings and we discuss  low
energy constraints coming from  oblique radiative
corrections. Section \ref{lasersec} contains a brief
description and expressions for the laser
backscattering mechanism as well as the discussion about the
cross sections. Our results for several  relevant kinematical
distributions of the final state particles are shown in
Sec.\ \ref{analsec}, and we summarize our conclusions in
Sec.\ \ref{conclusec}.

\section{Anomalous Quartic Couplings: Low Energy Constraints}
\label{coupsec}

In this work, we study some operators that generate anomalous quartic
vector-boson vertices. We consider genuinely quartic operators,
that is, operators which do not induce new trilinear vertices.  With
this choice, the strength of the quartic vertices cannot be constrained,
for instance, by LEP II bounds on the anomalous trilinear vertices.
Furthermore, since we are interested in probing anomalous couplings in
an $e \gamma$ collider, we concentrate on operators that involve at
least one photon. 

In constructing effective operators associated to such
anomalous couplings we employ the formalism of Ref.\
\cite{schild}. We  require the existence of a custodial
$SU(2)_{WI}$ symmetry,   which  forbids any contribution to
the $\rho$ parameter. We must also require that the
phenomenological lagrangians are invariant under local
$U(1)_{\text em}$ symmetry. The lowest order operators
that comply with the above requirements and give genuinely
quartic couplings, involving at least one photon, are of
dimension six. If we restrict our analyses to $C$ and $P$
conserving interactions, it is easy to see that there are
two independent operators involving two photons
\cite{cana1}
\begin{eqnarray} 
{\cal L}_0 & = & - \frac{\pi \alpha}{4 \Lambda^2}\,
a_0\, F^{\mu\nu}F_{\mu\nu}  W^{(i)\alpha} W^{(i)}_\alpha  \; , 
\label{lag:0} \\ 
{\cal L}_c & = & - \frac{\pi \alpha}{4 \Lambda^2}\,
a_c\, F^{\mu\alpha}  F_{\mu\beta} W^{(i)}_\alpha W^{(i)\beta} \; , 
\label{lag:c} 
\end{eqnarray} 
and one operator exhibiting just one photon
\begin{equation} 
{\cal L}_n  =  i \frac{\pi \alpha}{4 \Lambda^2}\, a_n\,
 \epsilon_{ijk} W^{(i)}_{\mu\alpha} W^{(j)}_\nu W^{(k)\alpha} F^{\mu\nu}
\; ,  
\label{lag:new} 
\end{equation}  
where $W^{(i)}$ is the $SU(2)_{WI}$ triplet and $F^{\mu\nu}$
is the  electromagnetic field strength.  When written in
terms of the physical fields $W^+$, $W^-$, and  $Z^0=W^3
\cos \theta_W$,  the effective lagrangians
(\ref{lag:0},\ref{lag:c}) give rise to anomalous
$W^+W^-\gamma\gamma$ and $Z^0Z^0\gamma\gamma$ couplings
while (\ref{lag:new}) generates a new $W^+W^-Z^0\gamma$
vertex. The $Z^0Z^0\gamma\gamma$ vertex is particularly
interesting since it is absent in the SM. 

It is important to analyze the possible low-energy
constraints on these lagrangians due to their one-loop
contribution to the vacuum polarization diagrams (oblique
corrections). In principle, ${\cal L}_n$ contributes only to
the $Z\gamma$ two-point function at the one-loop level, and
in consequence to $\sin \overline \theta_W$, whose
definition is based on the $Z^0$ asymmetries. However, such
contribution is proportional to the $W$ momentum in
the loop, and consequently the loop integral vanishes.
Therefore there are no low-energy constraints on ${\cal
L}_n$.

The lagrangians ${\cal L}_0$ and ${\cal L}_c$ contribute to
the photon, $W$, and $Z$ two point functions. Due to the
structure of the lagrangians, the contributions to the $W$
and $Z$ self-energies are constant, {\em i.e.\/} they do not
dependent on the external momentum, and are related by the
$SU(2)_{WI}$ custodial symmetry 
\begin{equation} 
{\Pi_{WW}}_{(0,c)} (q^2) = c_w^2 {\Pi_{ZZ}}_{(0,c)}(q^2) \; ,
\end{equation}
where $c_w = \cos \theta_W$. Therefore, as expected, they do not
contribute to the $\rho$ parameter, and also their contribution to $\sin
\overline \theta_W$ vanishes. 

Since these lagrangians preserve the local $U(1)_{\text em}$
symmetry, the photon self-energy has the form
\begin{equation}
{\Pi_{\gamma \gamma}}_{(0,c)} (q^2)=q^2 {\Pi'_{\gamma \gamma}}_{(0,c)} \; ,
\end{equation}	
where ${\Pi'_{\gamma \gamma}}_{(0,c)}$ is constant due to the explicit
form of the anomalous interactions.  Therefore, there is no contribution
to the running of the electromagnetic coupling either.

However, ${\cal L}_0$ and ${\cal L}_c$ affect the value of $\Delta r$, meaning
that the $S$ and $U$ parameters of Ref.\ \cite{Takeuchi,Hagiwara} or,
equivalently, $\epsilon_2$ and $\epsilon_3$ in notation of Ref.\
\cite{Altarelli} are modified.  In order to estimate this effect, we evaluate
${\Pi'_{\gamma \gamma}}_{(0,c)}$ using a gauge-invariant cutoff regularization
\cite{Zeppenfeld} 
\begin{eqnarray*}
{\Pi'_{\gamma \gamma}}_{0} = \frac{2\pi\alpha a_0}{\Lambda^2}
\biggl[ &-&\frac{M_W^2}{8\pi^2}\left(1+\frac{1}{2c_w^4}\right)+
 \frac{3 M_W^2}{16\pi^2}\ln\frac{\Lambda^2}{M_W^2}+\frac{3 M_W^2}{32\pi^2c_w^4}
\ln\frac{\Lambda^2 c_w^2}{M_W^2}-
\frac{\Lambda^2}{16\pi^2}\left(1+\frac{1}{2c_w^2}\right)\biggr] \; ,\\
{\Pi'_{\gamma \gamma}}_{c} = \frac{2\pi\alpha a_c}{\Lambda^2}
\biggl[&-&\frac{M_W^2}{128\pi^2}\left(1+\frac{1}{2c_w^4}\right)+
 \frac{3 M_W^2}{64\pi^2}\ln\frac{\Lambda^2}{M_W^2}+\frac{3 M_W^2}{128\pi^2c_w^4}
\ln\frac{\Lambda^2 c_w^2}{M_W^2}  \\ 
&-& \frac{\Lambda^2}{32\pi^2}\left(1+\frac{1}{2c_w^2}\right)-
\frac{3\Lambda^4}{64\pi^2 M_W^2}\biggr] \; .
\end{eqnarray*}

The cutoff $\Lambda$ can be  identified with the scale of the new
physics appearing in the lagrangian, whereas the coefficients $a_0$ and
$a_c$ remain arbitrary. In the following, we choose for the sake of
definiteness $\Lambda=M_W$.  Using the definitions of S and U of Ref.\
\cite{Hagiwara}, which for the new contributions reduce to the original
definitions of Ref.\ \cite{Takeuchi}, we obtain
\begin{eqnarray} 
(\Delta S)_0=0.17~ a_0  \;\;\;\;\; , \;\;\;\;\; (\Delta U)_0=0.051~ a_0  \; ,
\nonumber \\
(\Delta S)_c=0.055~ a_c \;\;\;\;\; , \;\;\;\;\; (\Delta U)_c=0.016~ a_c   \; .
\label{teor}
\end{eqnarray}

The present experimental values for the S and U parameters can be found
in Ref.\ \cite{Hagiwara} while the SM contributions can be found in
Ref.\ \cite{Zeppenfeld}. Assuming $m_{\text top} =
130$ GeV and $m_{H} = 1000$ GeV, the one sigma result for
$\Delta S$ is
\begin{equation}
-0.77 < \Delta S < 0.11 \; .
\label{exp}
\end{equation}
Comparing the theoretical results (\ref{teor}) with the
experimental ones (\ref{exp}), we obtain, at the one
sigma level, the following constraints.
\begin{eqnarray*}
-4.5  <  a_0  <  0.64 \\
-11  <  a_c  <  5.8 
\end{eqnarray*}

As we show in this paper, the bounds on $a_{0,c}$ coming
from low energy data are one to two orders of magnitude
(depending on the top and Higgs masses) weaker than the
accelerator limits that are obtained via the study of vector
boson production in $e\gamma$ collisions.

%******************************************************************************

\section{Laser Backscattering and Cross Section}
\label{lasersec}

The most promising mechanism to generate a hard photon beam
in an  $e^+e^-$ linear collider is the laser backscattering.
When soft photons from a few eV laser collide with an
electron or positron beam, a large flux of photons, carrying
a great amount of the parent fermion energy, is generated. The
spectrum of the laser backscattered photons is 
\cite{telnov}
\begin{equation}
F_L (x) \equiv \frac{1}{\sigma_c} \frac{d\sigma_c}{dx} = 
\frac{1}{D(\xi)} \left[ 1 - x + \frac{1}{1-x} - \frac{4x}{\xi (1-x)} +  
\frac{4
x^2}{\xi^2 (1-x)^2}  \right] \; ,
\label{f:l}
\end{equation}
with
\[
D(\xi) = \left(1 - \frac{4}{\xi} - \frac{8}{\xi^2}  \right) \ln (1 + \xi) +
\frac{1}{2} + \frac{8}{\xi} - \frac{1}{2(1 + \xi)^2} \; ,
\]
where $\sigma_c$ is the Compton cross section, $\xi \simeq 4
E\omega_0/m_e^2$,  $m_e$ and $E$ are the electron mass and
energy respectively, and $\omega_0$ is the laser photon
energy. The fraction $x$ represents the ratio between the
scattered photon and initial electron energy for  the
backscattered photons traveling along the initial electron
direction. The maximum possible value of $x$ is 
\begin{equation}
x_{\text max} = \frac{\omega_{\text max}}{E} 
= \frac{\xi}{1+\xi} \; ,
\end{equation}
with $\omega_{\text max}$ being the maximum scattered
photon energy. 

It is interesting to notice that $x_{\text max}$ grows
with the laser and the electron energy.  However, $\omega_0$
should be less than $m_e^2/ \omega_{\text max}$ in order
to avoid that the interaction of laser photons and
backscattered ones create a pair $e^+e^-$, which would
reduce the conversion of electrons to photons.  In our
calculation we have chosen the laser energy in order to
maximize the backscattered photon energy without spoiling
the luminosity.  This can be accomplished by taking $\xi=
2(1+\sqrt{2}) \simeq 4.8$. With this choice, the photon
spectrum exhibits a peak close to its maximum which occurs
at $x_{\text max} \simeq 0.83$. 

The total cross sections are obtained by folding the elementary cross sections
with the photon distribution 
\begin{equation}
\sigma(s)=\int_{0}^{x_{\text max}} dx ~ F_L (x) ~ 
\hat\sigma_{VV^\prime F} (\hat s) \; ,
\end{equation}
where $\hat\sigma_{VV^\prime F}$ are the cross sections,
evaluated at $\hat s = xs$, for the reactions $e\gamma
\rightarrow VV^\prime F$,  with $V, V^\prime =W$, $Z$, or
$\gamma$ and $F=e$ or $\nu$. 

The  analytical evaluation of the cross section $\hat\sigma_{VV^\prime
F}$  for each of the  subprocesses (\ref{wwe}--\ref{wzn}) is very
lengthy and tedious despite being straightforward. In order to perform
these calculations in a efficient way, we used a numerical method 
similar to the one described in Ref.\ \cite{barger}.   We  verified
explicitly that our amplitudes are Lorentz invariant and
$U(1)_{\text em}$
gauge invariant.   The numerical integrations were performed using a
Monte Carlo routine \cite{lepage}, and we  tested our results against
possible statistical fluctuations.

%**************************************************************

\section{Double Vector Boson Production: Distributions and Results}
\label{analsec}

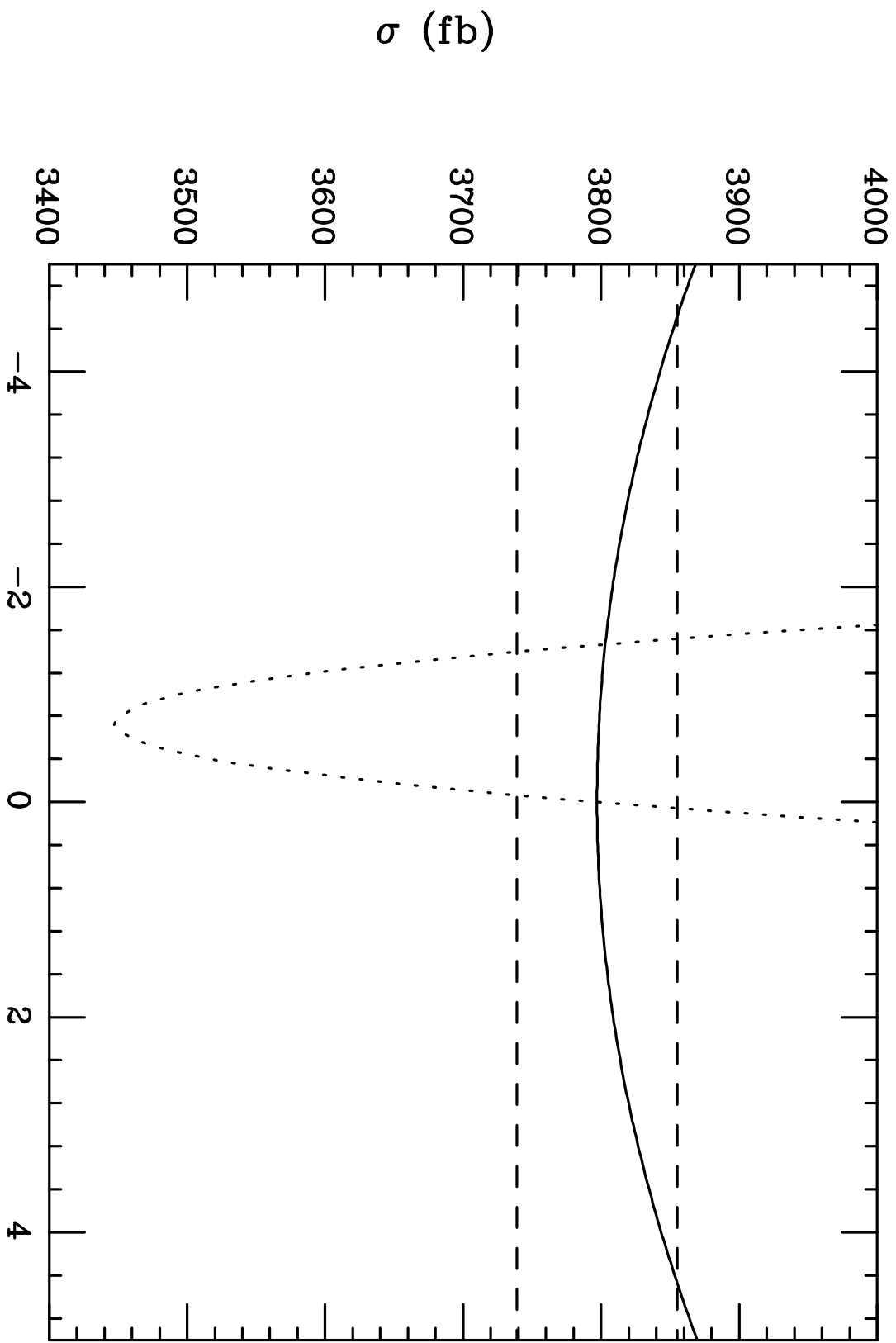
\begin{figure}
\protect
\caption{Total cross section for the process $WWE$  as a function
of the anomalous couplings $a_c$ (dotted line) and $a_n$ (solid line). 
The $3\sigma$ interval around the SM value of the cross section  is
marked by horizontal dashed lines. We assumed an integrated luminosity
${\cal L}=10$ fb$^{-1}$ for the $e^+e^-$ collider.
\label{sig:cn}}
\end{figure}

Now we have all the  necessary ingredients to perform a
detailed study of the production of two gauge boson in
$e\gamma$ colliders, taking into account anomalous quartic
couplings. First of all, we analyzed the effect of the
anomalous couplings on the total cross section of the
processes (\ref{wwe}--\ref{wzn}). The anomalous couplings
$a_0$ and $a_c$ contribute to the processes $WWE$, $ZGE$,
$ZZE$,  and $WGN$ while the coupling $a_n$ participate only
of the processes  $WWE$ and $WZN$.  The anomalous cross
sections for these processes are quadratic functions of the
parameters $a_i$ ($i=0, ~c, ~n$) {\em i.e.}
\begin{equation}
\sigma_{\text AN}~ =~ \sigma_{\text SM}~ +~ a_i~ \sigma_{\text int}^i~  +~ 
a_i^2~ \sigma_{\text ano}^i \; ,
\label{param}
\end{equation}
where $\sigma_{\text SM}$ is the Standard Model cross section and
$\sigma_{\text int}$ ($\sigma_{\text ano}$) is the interference (pure
anomalous) contribution.  We present in Table \ref{tot:cs}
the values of $\sigma_{\text SM,int,ano}$ for the interactions
\ref{lag:0}--\ref{lag:new}, assuming that only one anomalous
coupling is different from zero at one time. For events
containing a photon in the final state, we imposed a cut on
the photon transverse momentum $p_T^\gamma>15$ GeV, which
not only guarantees that our results are free of infrared
divergences but also mimics the performance of a typical
electromagnetic calorimeter.  As an illustration of the
typical behaviour of $\sigma_{\text AN}$, we exhibit in Fig.\
\ref{sig:cn}  the total cross section for the process $WWE$
as a function of $a_c$ and $a_n$.

From the Table \ref{tot:cs} we  learn that $WWE$ and $ZZE$
are the most sensitive processes to $a_0$ and $a_c$. In the
$WWE$ case, the SM cross section is larger than the crossed
channel $e^+e^- \rightarrow W^+ W^- \gamma$ by more than one
order of magnitude \cite{cana1}, due to the $t$-channel
photon exchange diagrams \cite{effe} that dominates in the
former case. These  diagrams involves the anomalous quartic
coupling making this process very sensitive to $a_0$ and
$a_c$.  For the process $ZZE$, the SM calculation does not
contain a $t$-channel photon exchange. However, the
introduction of the anomalous couplings $Z^0Z^0\gamma\gamma$
gives rise to this kind of contribution,  making $ZZE$ very
sensitive to the anomalous couplings, despite the small
value of its SM cross section. We should notice that for the
process $ZGE$, the $a_{0,c}$ anomalous coupling also
introduces a new contribution, due to the vertex
$Z^0Z^0\gamma\gamma$, that has a $Z$ exchange in the
$t$-channel. Nevertheless, in the effective $Z$
approximation \cite{eff:w}, we note that the enhancement is
at most proportional to $\ln(s/M_Z^2)$ for the exchange of
transverse $Z$'s, and consequently the $ZGE$ becomes almost
insensitive to the anomalous couplings.

The $WWE$ production turned out to be rather insensitive to the
anomalous coupling $a_n$ since the anomalous vertex $W^+W^-Z^0\gamma$
does not contribute to the dominant diagrams of this process, which are
the ones involving the exchange of a photon in the $t$-channel.  In
fact, in the effective $Z$ and $\gamma$ approximations we can see that
the anomalous contribution is suppressed by a factor $\ln(s/M_Z^2) /
\ln(s/m_e^2)$ in relation to the dominant terms. In spite of the cross
section for $WZN$ being smaller than the one for $WWE$, the process
$WZN$ is the most sensitive to this anomalous coupling.

In order to quantify the effect of the new couplings, we
define the statistical significance (${\cal S}$) of the
anomalous signal 
\begin{equation} 
{\cal S} = \frac{|\sigma_{\text AN} -
\sigma_{\text SM}|}{\sqrt{\sigma_{\text SM}}} \; 
\sqrt{\cal L} \; ,
\label{sig} 
\end{equation} 
which can be easily evaluated using the parametrization
(\ref{param}) with the coefficients given in Table
\ref{tot:cs}.  We list in Table \ref{tab:sig} the values of
the anomalous couplings that correspond to a $3\sigma$
effect for the different processes, assuming an integrated
luminosity ${\cal L}= 10$  fb$^{-1}$ for the associated
$e^+e^-$ collider.  We checked that our results do not
change by more than a factor of $2$  if we adopt realistic
values for the efficiency of the vector boson reconstruction
and increase the  value of ${\cal S}$ to 5.

We can see from Table \ref{tab:sig} that the $3\sigma$ limits for
the $a_0$ and $a_6$ couplings are approximately one order magnitude
better than the limits one can get in the $e^+ e^-$ mode
\cite{cana1,numbers}.  Our limits are of the same order than the ones
obtained through the reaction $\gamma \gamma \rightarrow W^+W^-$ with
unpolarized beams \cite{cana2,numbers} and are within a factor of $5$ of
the ones from the reaction $\gamma\gamma \rightarrow Z^0 Z^0$
\cite{cana2,numbers}.  On the other hand, we must stress that the
anomalous $a_n$ vertex can not be studied so well in the $\gamma\gamma$
mode of the collider since it contributes only to processes like
$\gamma\gamma \rightarrow W^+W^-Z^0$ which have a smaller phase space.
From what we learnt above, we also expect that the attainable limits at
$e^+e^-$ to be much worse than the ones from $\gamma\gamma$ and
$e\gamma$.

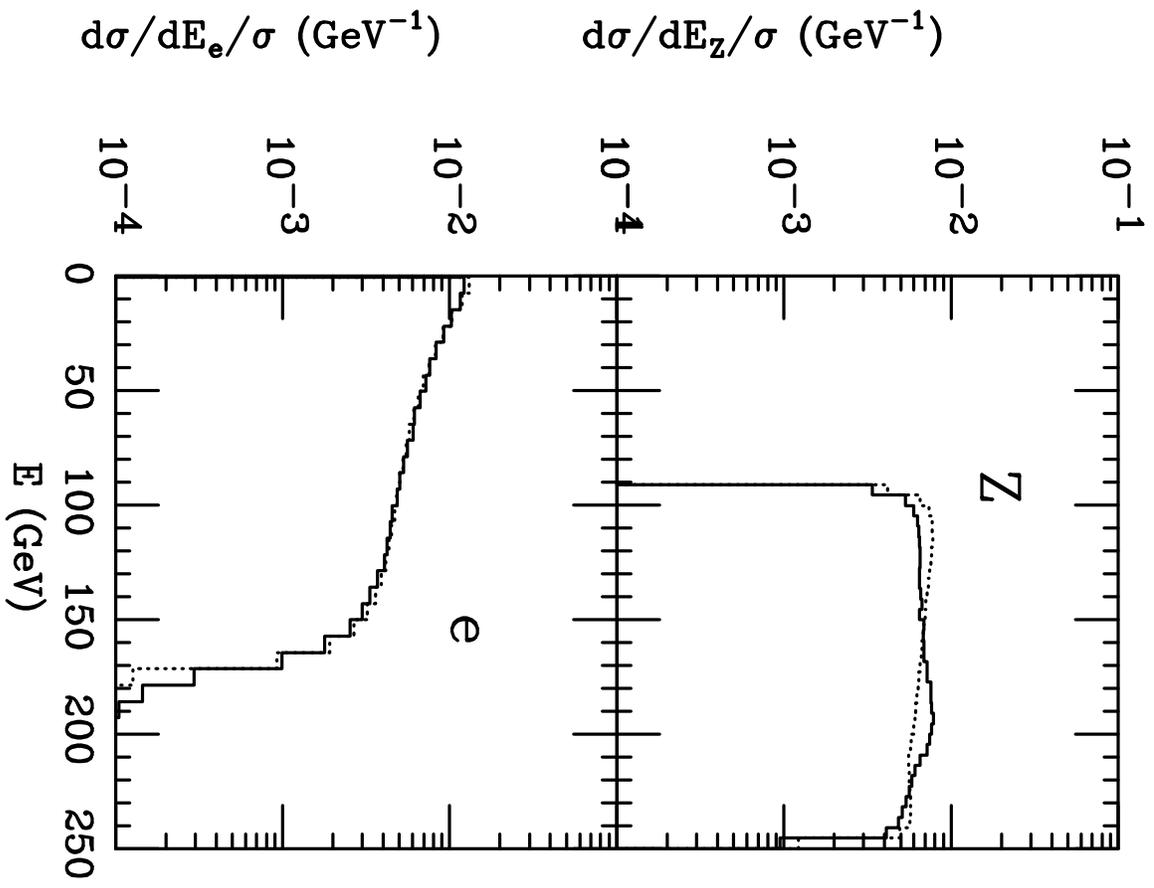
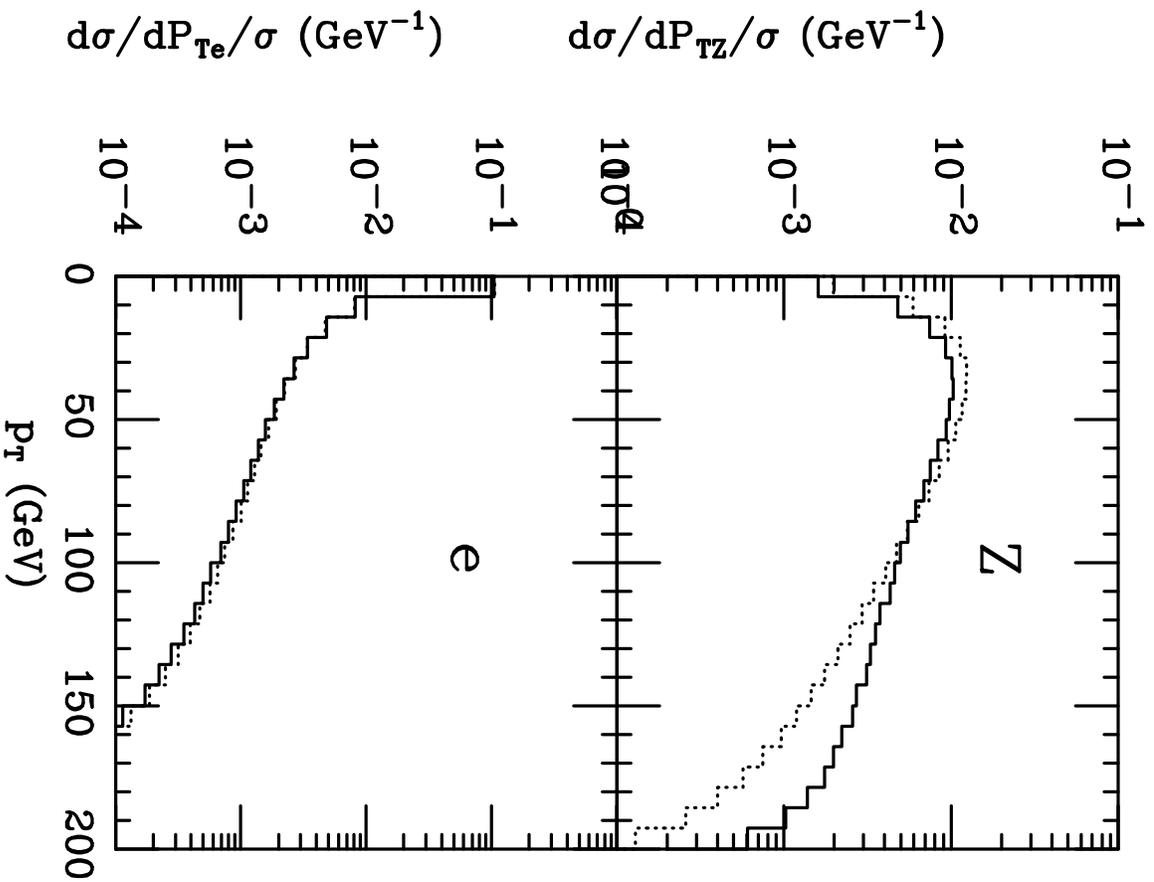
\begin{figure}
\protect
\caption{Energy and transverse momentum distributions for $Z$ and $e$ in the
process $ZZE$. The dotted line is the SM prediction and the solid line
is the anomalous distribution for the positive $3\sigma$ limit
$a_0=0.028$. 
\label{zze:ept}}
\end{figure}

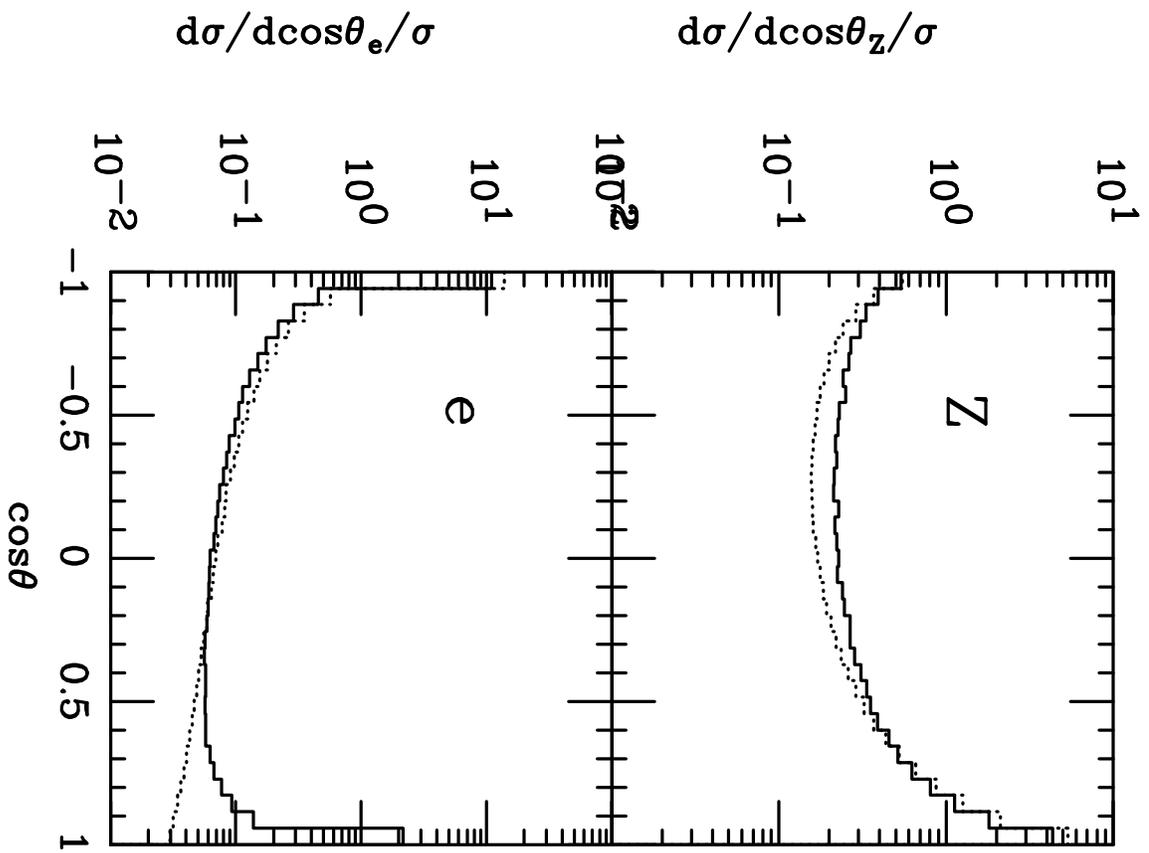
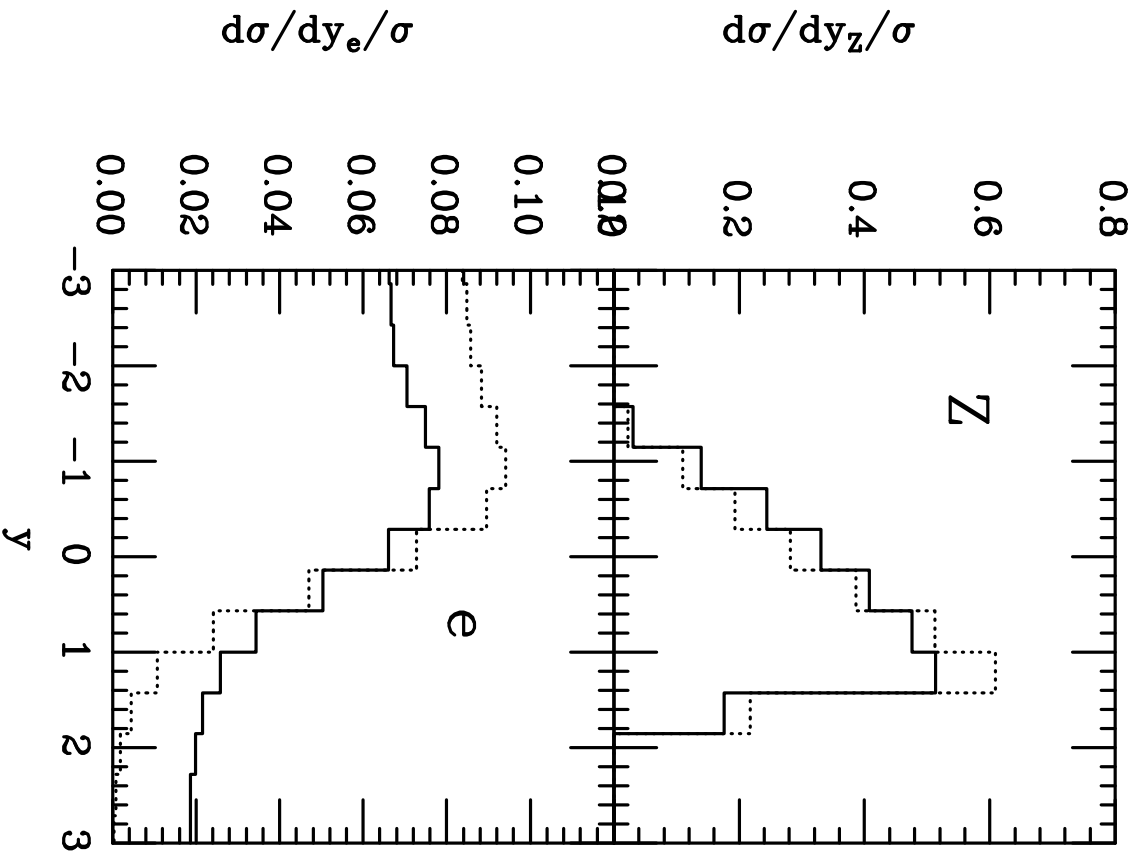
\begin{figure}
\protect
\caption{Angle with the beam pipe and rapidity distributions
for $Z$ and $e$  in the process $ZZE$. The conventions are
the same as  in Fig.\ \protect\ref{zze:ept}.
\label{zze:cosy}}
\end{figure}

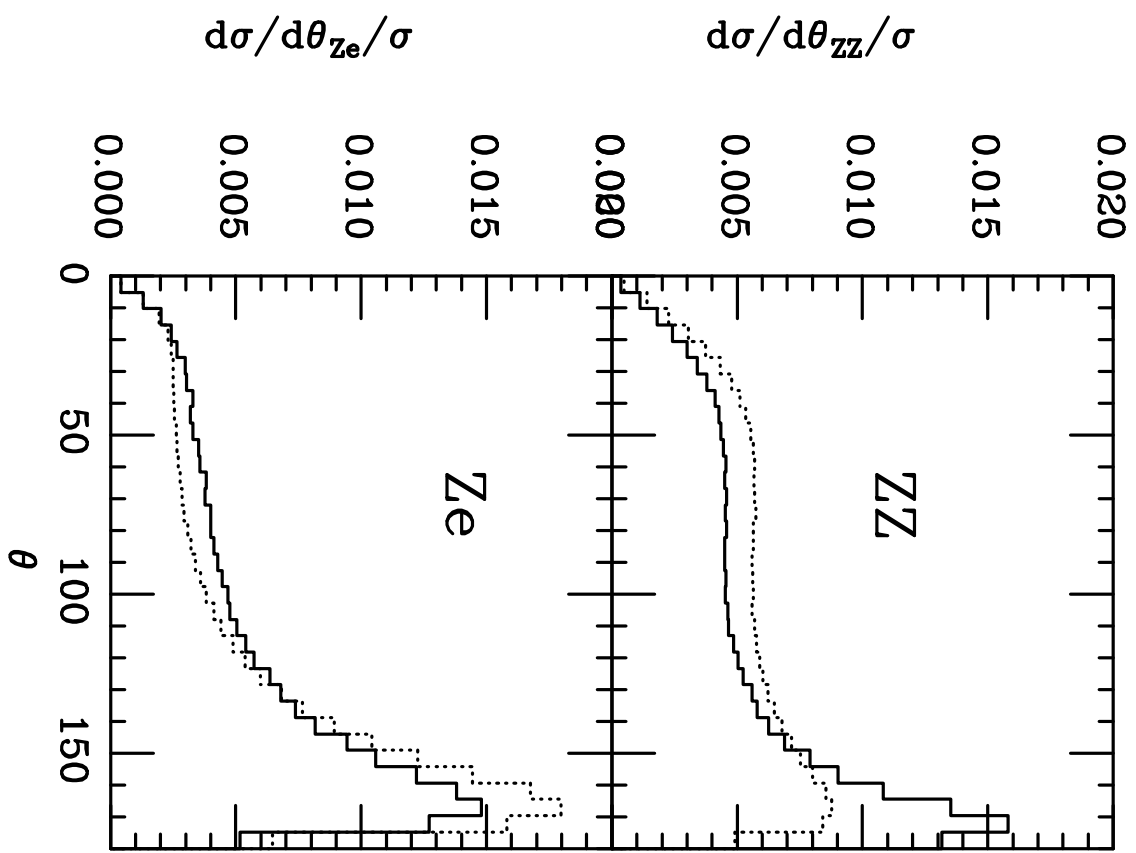
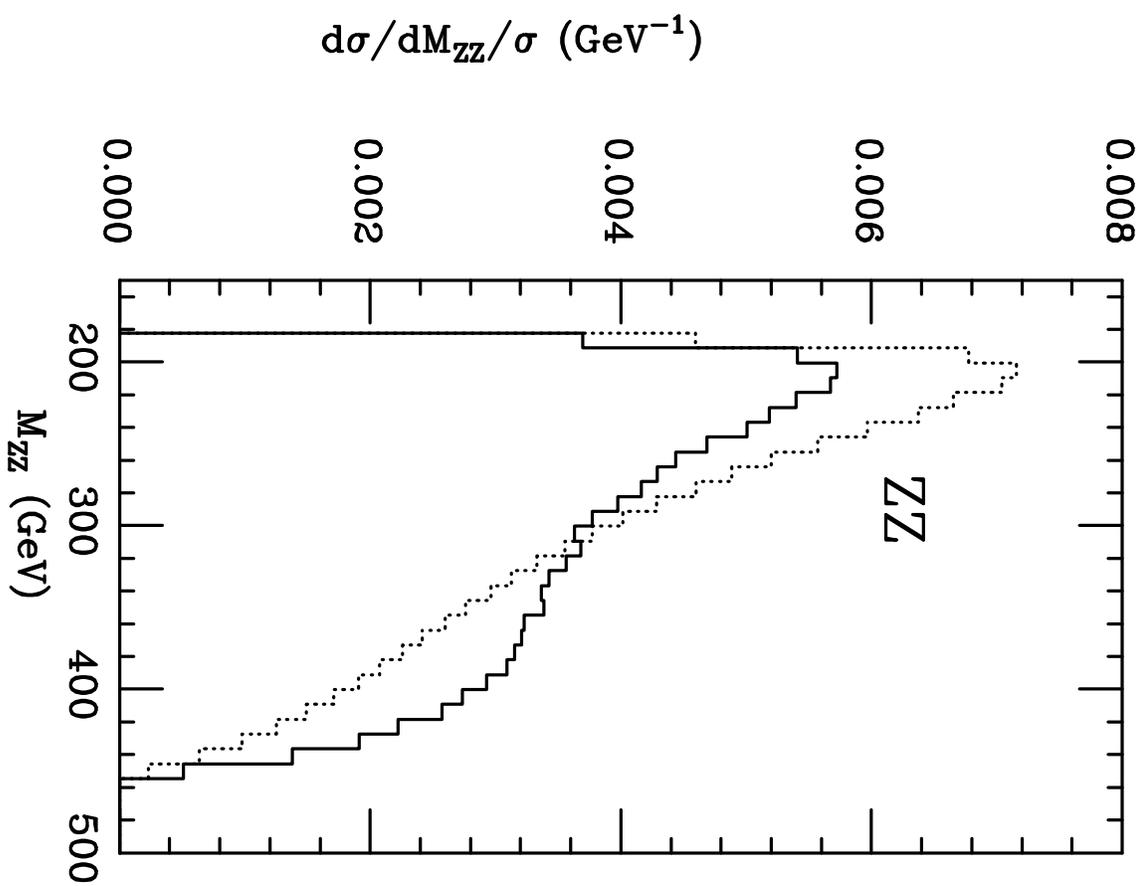
\begin{figure}
\protect
\caption{Distribution in the angle between the final
particles and invariant mass distribution of the $ZZ$ pair
in the process $ZZE$. The conventions are the same as  in
Fig.\ \protect\ref{zze:ept}.
\label{zze:angm}}
\end{figure}

The study of the anomalous couplings through the total cross section
of the different processes is the crudest thing that can be done. In
principle, the shapes and values of the kinematical distributions can
be used to increase the sensitivity to the anomalous couplings,
improving consequently the bounds on the new couplings and yielding
further information.  In Figs.\ \ref{zze:ept} to \ref{zze:angm} we
compare some SM distributions for the $ZZE$ process with the ones
corresponding to the anomalous coupling $a_0=0.028$, which gives a
$3\sigma$ deviation in the total cross section.  From these figures,
we can see that the effect of the anomalous interactions in the
distributions is more than a mere overall normalization.  For
instance, Fig.\ \ref{zze:ept} exhibits the energy and transverse
momentum distributions for $Z$ and $e$, and shows that the existence of
an anomalous coupling favors the production of higher $p_T$ vector
bosons.

More dramatic is the difference in the angular (or rapidity)
distribution of the final particles, as seen in Fig.\
\ref{zze:cosy}. In the SM, the final electron goes
preferentially in the opposite direction of the initial
electron, as in a ``Compton-like'' process. However, the
anomalous $a_{0}$ coupling contributes with a new photon
$t$-channel, which is dominated by low transfer momenta.
This effect makes the electron go mainly in the forward
direction and leads to an electron angular distribution
which is very distinct from the one expected in the SM.  The
distributions in the angle between the final particles and
the invariant-mass distribution of the $ZZ$ pair are shown
in Fig.\ \ref{zze:angm}. We see that the effect of the
anomalous coupling is to increase the average angle between
the $Z$'s while, at the same time, to reduce the average
angle between the $Z$ and the electron.  Furthermore, the
anomalous interaction also favors higher invariant masses
for the $ZZ$ pairs,  since the new couplings are
proportional to the photon momentum, which are very large in
the case of backscattered photons.

Despite the large sensitivity of the $ZZE$ process to the anomalous
couplings $a_{0,c}$, we verified that the anomalous $3\sigma$
distributions for the couplings $a_0$ and $a_c$ are indistinguishable,
since all the cases give rise to distributions that exhibit the same
general behaviour. Therefore, the analysis of the process $ZZE$ can
only indicate the existence of anomalous couplings, but not their
origins and explicit form.

We also verified that in the case of the $WWE$ process, the
deviations from the SM predictions for the kinematical
distributions is equivalent to the $ZZE$ process. In spite
of giving looser bounds on the anomalous couplings, the
$WWE$ process has a large cross section which can give rise
to a substantial statistic.

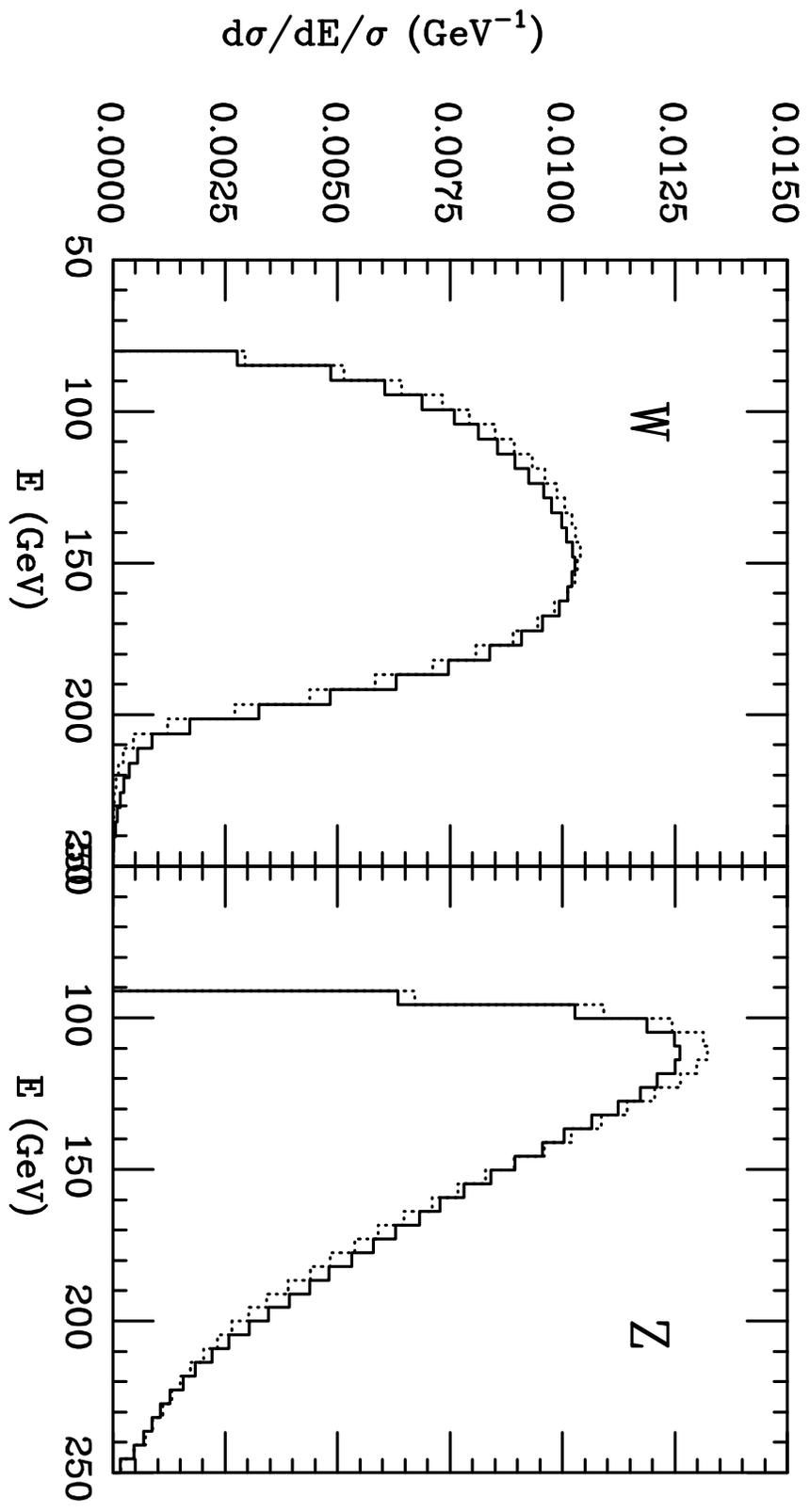
\begin{figure}
\protect
\caption{Energy distribution for $W$ and $Z$ in the process
$WZN$. The dotted line is the SM prediction and the solid
line is the anomalous distribution for the positive
$3\sigma$ limit $a_n=0.74$. 
\label{wzn:e}}
\end{figure}

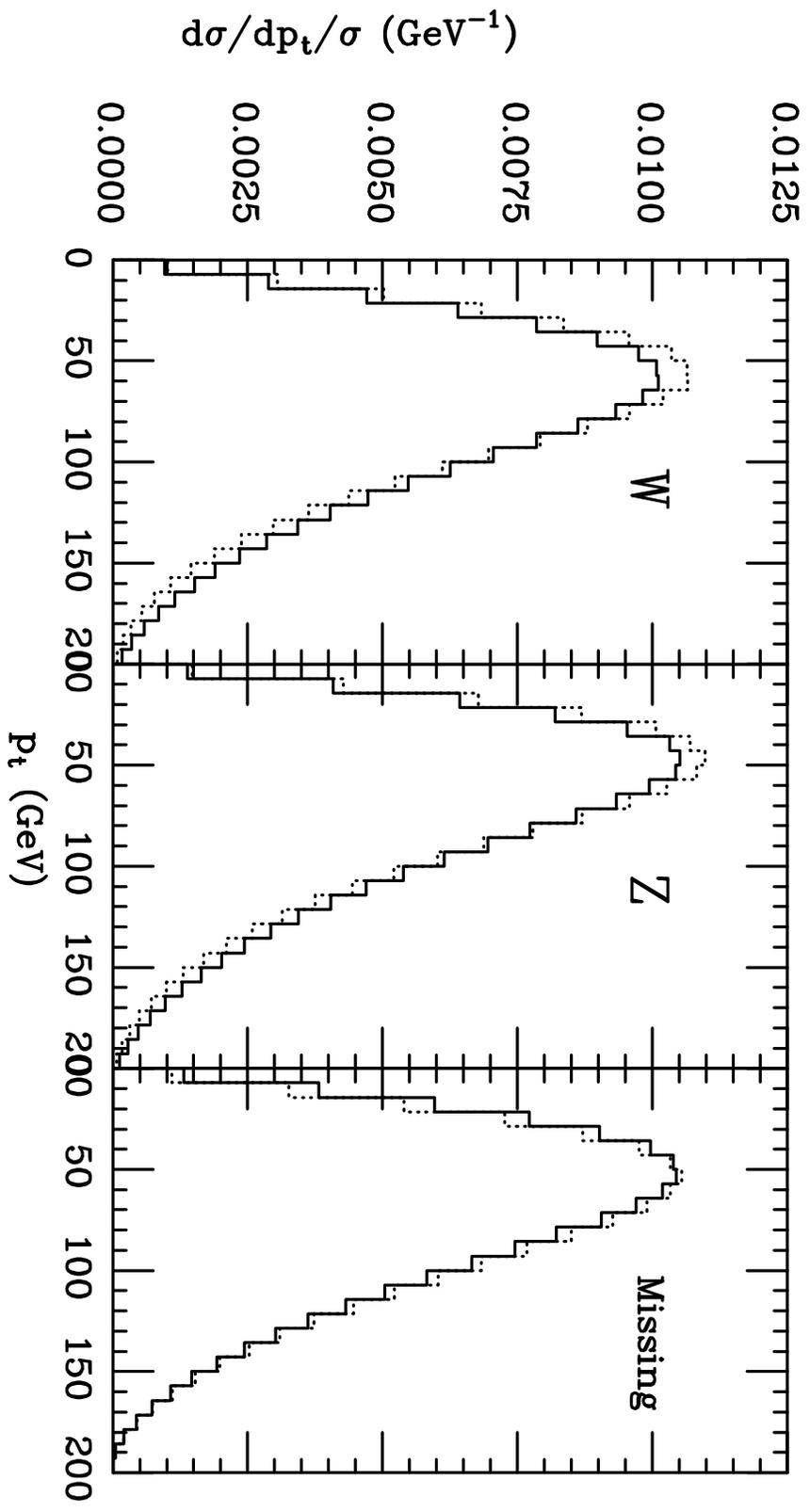
\begin{figure}
\protect
\caption{Transverse momentum distribution for $W$, $Z$ and
missing (neutrino) in the process $WZN$. The conventions are
the same as  in Fig.\ \protect\ref{wzn:e}.
\label{wzn:pt}}
\end{figure}

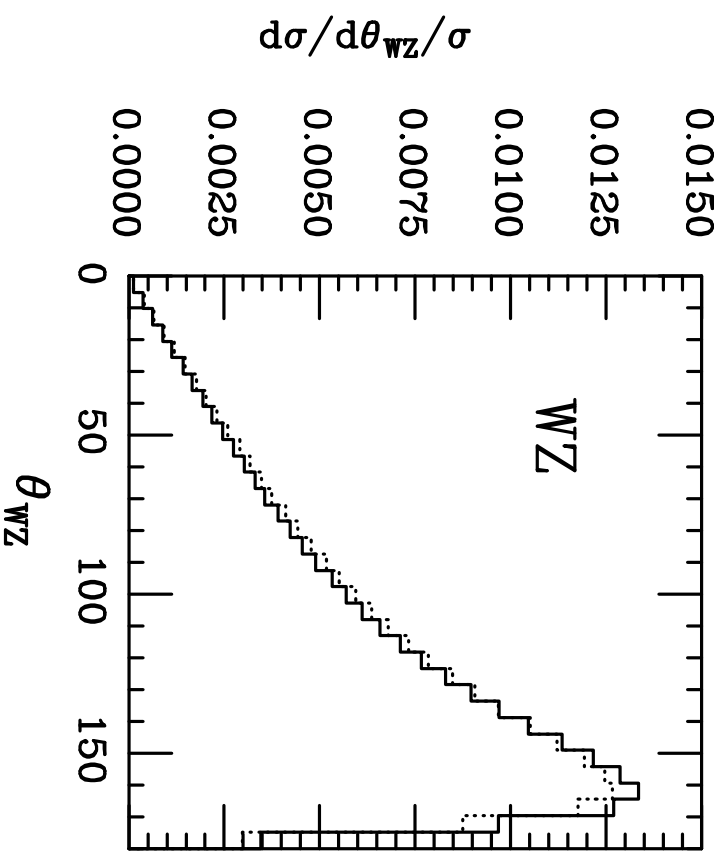
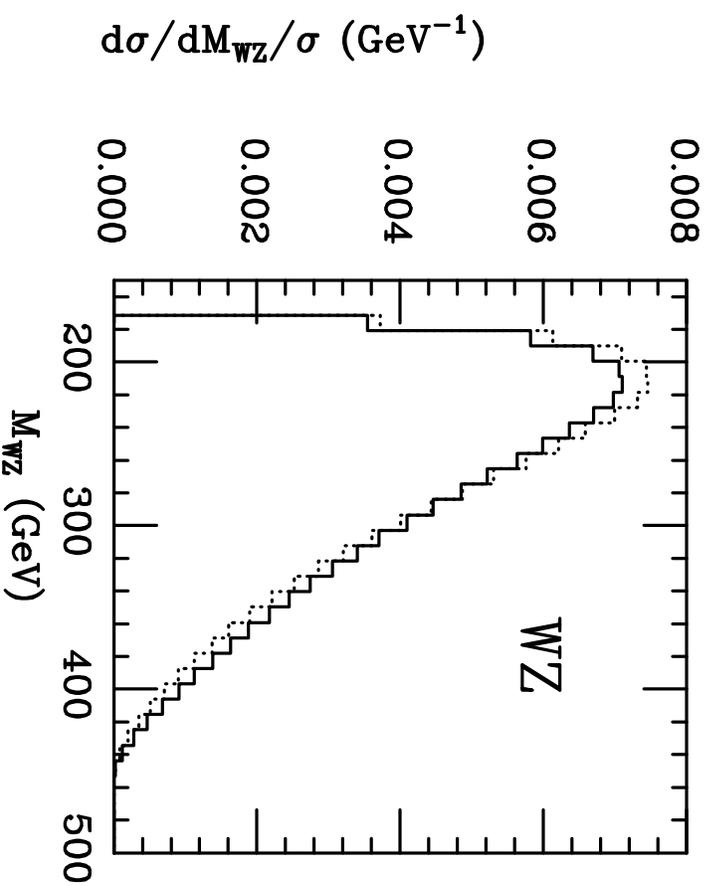
\begin{figure}
\protect
\caption{Distribution of the angle between the gauge bosons
and invariant mass distribution of the gauge bosons in the
process $WZN$. The conventions are the same as  in Fig.\
\protect\ref{wzn:e}. 
\label{wzn:angm}}
\end{figure}

In Figs.\ \ref{wzn:e}-\ref{wzn:angm} we exhibit some
distributions for the $WZN$ process for the SM and also
including the anomalous coupling $a_n=0.74$, which
corresponds to a $3\sigma$ deviation in the total cross
section. We verified that, in this process, the behaviour of
the anomalous distributions for the $3\sigma$ limits
($a_n=0.74$ and $a_n=-1.2$) are basically the same, and
consequently they are indistinguishable.  One interesting
property of the $a_n$ anomalous vertex is that its intensity
is linearly related to the photon momentum, and consequently
its contribution to the processes is more important for
highly energetic photons, whose energy is close to the
maximum of the laser backscattering spectrum. Therefore,
the presence of the anomalous vertex increases the available
energy for the reaction causing the distribution of momenta
and energy of the final gauge bosons to be shifted towards
high energy (see Figs.\ \ref{wzn:e} and \ref{wzn:pt}).  This
argument allows us to conclude that the invariant mass
distributions of the vector bosons are also increased at
high invariant masses, as can be seen from Fig.\
\ref{wzn:angm}.

\section{Summary and Conclusions}
\label{conclusec}

In this paper we studied the capability of an $e^+e^-$
collider, operating in the $e\gamma$ mode to search for
anomalous four-gauge-boson interactions.  We concentrated
on those that contain at least one photon and that
do not exhibit a trilinear coupling associated to the it,
and required the anomalous lagrangians to be invariant under
a custodial $SU(2)_{WI}$ and under local $U(1)_{\text em}$ 
symmetries.  One interesting property of the
interactions we analyzed is that the available low energy
constraints on these anomalous couplings are very loose.

The search for anomalous couplings has been performed
through the analyses of production of gauge boson pairs,
$e\gamma \rightarrow VV^\prime F$ ($V, V^\prime =W$, $Z$, or
$\gamma$ and $F=e$ or $\nu$).   Our results show that $ZZE$
and $WWE$ are the most sensitive processes to the anomalous
couplings involving two photons, $a_0$ and $a_6$. The
$3\sigma$ limits on these couplings obtained from the study
of the total cross section for the $VV^\prime F$ production
are approximately one order magnitude better than the limits
using the $e^+ e^-$ mode \cite{cana1} and a factor of 5
worse than the limits coming from the $\gamma\gamma$ mode
\cite{cana2}. The anomalous coupling $a_n$, which only
involves one photon in the quartic vertex, cannot be
efficiently studied in the $\gamma\gamma$ or $e^+ e^-$ modes
of the collider, being $WZN$ the process most sensitive to
it.

The bottom line of our work is that an $e^+e^-$ collider
operating in the $e\gamma$ mode will be able to increase
substantially the potential of analyses of anomalous
four-gauge-boson interactions with respect to the $e^+e^-$
mode. Despite of the $e\gamma$ mode being more sensitive to
the presence of further contributions to the standard model,
it will be a hard task to discriminate the form and origin
of the new contributions.

%******************************************************************************

\acknowledgments

We would like to thank F.\ Halzen and D.\ Zeppenfeld for valuable
discussions.  S.F.N. is very grateful to the Institute for Elementary
Particle Physics Research of the Physics Department, University of
Wisconsin -- Madison for their kind hospitality.  This work was
supported by the University of Wisconsin Research Committee with funds
granted by the Wisconsin Alumni Research Foundation, by the U.S.\
Department of Energy under Contract No.~DE-AC02-76ER00881, by the
Texas National Research Laboratory Commission under Grant
No.~RGFY9273, by Conselho Nacional de Desenvolvimento Cient\'{\i}fico
e Tecnol\'ogico (CNPq-Brazil), and by the National Science Foundation
under Contract INT 916182.

\appendix
\section*{Feynman Rules}

We present in this appendix the Feynman rules for the anomalous
interactions ${\cal L}_0$ (Eq.\ \ref{lag:0}), ${\cal L}_c$ (Eq.\ \ref{lag:c}),
and ${\cal L}_n$ (Eq.\ \ref{lag:new}). Our conventions are that all the
momenta are incoming to the vertex. The interactions ${\cal L}_0$ and
${\cal L}_c$ give rise to anomalous contributions to $W^{+
\; \mu} (p_+) - W^{- \; \nu} (p_-) - \gamma^\alpha (p_1) -\gamma^\beta
(p_2)$ given by
\begin{equation}
i \frac{2 \pi \alpha}{\Lambda^2} ~  a_0 ~  g_{\mu\nu} ~ 
\left[ g_{\alpha\beta} ~ (p_1.p_2) - p_{2  \alpha} p_{1  \beta}  \right] \; ,
\label{rf:a0}
\end{equation}
and
\begin{eqnarray}
i \frac{\pi \alpha}{2 \Lambda^2} ~ a_c ~ \biggl[ &&
(p_1.p_2) \left(g_{\mu\alpha} ~  g_{\nu\beta} + 
g_{\mu\beta} ~ g_{\alpha\nu}\right)
+ g_{\alpha\beta} ~  \left( p_{1  \mu} p_{2  \nu} + 
p_{2  \mu} p_{1  \nu} \right) 
\nonumber \\
&& - p_{1  \beta} \left( g_{\alpha\mu} ~ p_{2  \nu} + 
g_{\alpha\nu} ~ p_{2  \mu}  \right)
- p_{2  \alpha} \left( g_{\beta\mu} ~ p_{1  \nu} +
g_{\beta\nu} ~  p_{1  \mu} \right)
\biggr] \; ,
\label{rf:ac}
\end{eqnarray}
respectively.  The contributions of these lagrangians to the anomalous
$Z^0-Z^0-\gamma-\gamma$ vertex is obtained by multiplying the above
results by $1/\cos^2\theta_W$ and making $W \rightarrow Z$.

The interaction ${\cal L}_n$ yields a contribution to $W^{+ \; \mu}
(p_+) - W^{- \; \nu} (p_-) - \gamma^\alpha (p_1) - Z^\beta (p_2)$ given
by
\begin{eqnarray}
i \frac{\pi \alpha}{4 \cos\theta_W \Lambda^2} ~ a_n ~ \biggl\{ &&
g_{\mu\beta} ~ \left[ g_{\nu\alpha} ~ p_1.(p_2 - p_+) -
p_{1  \nu} (p_2 - p_+)_\alpha \right] 
\nonumber \\
&& - g_{\nu\beta} ~ \left[ g_{\mu\alpha} ~  p_1.(p_2 - p_-) -
p_{1  \mu} (p_2 - p_-)_\alpha  \right]
\nonumber \\
&& + g_{\mu\nu} ~ \left[ g_{\alpha\beta} ~ p_1.(p_+ - p_- ) - 
( p_+ - p_- )_\alpha p_{1  \beta} \right] 
\nonumber \\
&&- p_{2  \mu} \left(g_{\alpha\nu} ~ p_{1  \beta} - 
g_{\alpha\beta} ~ p_{1  \nu}   \right)
+ p_{2  \nu} \left(g_{\alpha\mu} ~ p_{1  \beta} - 
g_{\alpha\beta} ~ p_{1  \mu}   \right)
\nonumber \\
&&- p_{-  \beta} \left(g_{\alpha\mu} ~ p_{1  \nu} - 
g_{\alpha\nu}  ~ p_{1  \mu}   \right)
+ p_{+  \beta} \left(g_{\alpha\nu} ~ p_{1  \mu} - 
g_{\alpha\mu} ~  p_{1  \nu}   \right)
\nonumber \\
&& - p_{+  \nu} \left(g_{\alpha\beta} ~ p_{1  \mu} - 
g_{\alpha\mu} ~ p_{1  \beta}   \right)
+ p_{-  \mu} \left(g_{\alpha\beta} ~ p_{1  \nu} - 
g_{\alpha\nu} ~ p_{1  \beta}  \right)
\biggr\} \; .
\label{rf:an}
\end{eqnarray}

%******************************************************************************
%**********
% TABLES
%**********
\widetext
\begin{table}
\caption{ Parameters characterizing the anomalous cross sections in fb for
the different processes and interactions. }
\label{tot:cs}
\begin{tabular}{c|c|c|c|c|c}
           &   $WWE$     &   $ZGE$    &   $ZZE$      &   $WGN$    &   $WZN$ \\
\tableline
$\sigma_{\text SM}$    & 3797. & 378. & 9.91 & 457. & 71.9 \\
\hline
$\sigma_{\text int}^0$ & 1361. & $-$0.0976 &  0.982 & 1.38 & --- \\
$\sigma_{\text ano}^0$ & 8005. & 20.7 & 3697. & 67.6 & --- \\
\hline
$\sigma_{\text int}^c$ & 959. & $-$0.673 &  0.960 & $-$17.4 & --- \\
$\sigma_{\text ano}^c$ & 657. & 9.65 & 318. & 31.6 & --- \\
\hline
$\sigma_{\text int}^n$ & 0.118 &  --- & --- & --- &  4.01 \\  
$\sigma_{\text ano}^n$ & 2.88  &  --- &  ---  &  --- & 9.22 \\
\end{tabular}
\end{table}

\begin{table}
\caption{Intervals for $a_0$, $a_c$, and $a_n$ 
corresponding to a deviation $\leq 3\sigma$ in the total cross
section. We also exhibit the difference ($\Delta\sigma$) between 
the anomalous cross
sections and the SM ones in fb for a $3\sigma$ effect.}
\label{tab:sig}
\begin{tabular}{c|c|c|c|c|c}
     &   $WWE$     &   $ZGE$    &   $ZZE$      &   $WGN$    &   $WZN$ \\
\tableline
$a_0$&($-$0.21, 0.036)&($-$0.94, 0.95)&($-$0.029, 0.028)&($-$0.56,
0.54) &  ---    \\
\hline
$a_c$&
$\begin{array}{c}
(-1.5,  -1.4) \\  
\mbox{and}  \\
  (-0.064, 0.059) \\
\end{array}$
 &($-$1.3,
1.4)& ($-$0.098, 0.095)&($-$0.57, 1.1)& ---\\
\hline
$a_n$&($-$4.5, 4.5)    &   ---      &    ---       &   ---  
&($-$1.2, 0.74)\\
\hline
$|\Delta\sigma|$ & 58. & 18. & 3.0 & 20. & 8.0 \\
\end{tabular}
\end{table}

%******************************************************************************

\protect

%***
\protect

%***
\protect

%***
\protect

%***
\protect

%***
\protect

%***
\protect


\begin{references}
\bibitem{numbers} The limits on $a_0$ and $a_6$ obtained from  $e^+ e^-$
collisions \cite{cana1} are $-0.4 \le a_0 \le 0.6 $ and $-0.7 \le a_c
\le 0.3$. The ones obtained through the reaction $\gamma \gamma
\rightarrow W^+W^-$ with unpolarized beams \cite{cana2} are $-6.0
\times 10^{-2} \le a_0 \le 3.0 \times 10^{-2}$ and $-0.1 \le a_c \le
3.5 \times 10^{-2}$, while the corresponding limits from $\gamma\gamma
\rightarrow Z^0 Z^0$ \cite{cana2} are $| a_0 | \le 4.0\times10^{-3} $ and 
$| a_c | \le 7.0 \times10^{-3}$.
\bibitem{effe} We checked that the total cross section for
the process $e + \gamma \rightarrow W^+ + W^- + e$ is very well
approximated by the reaction $e + \gamma \rightarrow e + \gamma +
(\gamma)_b \rightarrow W^+ + W^- + e$, with the bremsstrahlung photon
$ (\gamma)_b $ described by the equivalent photon 
(Weizs\"acker-Williams) approximation.
\end{references}
\end{document}